\documentclass[aps,pre,twocolumn,showpacs,superscriptaddress,preprintnumbers,amsmath,amssymb]{revtex4}

\usepackage{amsmath}
\usepackage{graphicx}
\usepackage{dcolumn}
\usepackage{bm}
\usepackage{float}

\begin{document}

\title{Shocks in unmagnetized plasma with a shear flow: Stability and magnetic field generation}

\author{M. E. Dieckmann}
\affiliation{Department of Science and Technology, Link\"oping University, SE-60174 Norrk\"oping, Sweden}

\author{A. Bock}
\affiliation{Department of Science and Technology, Link\"oping University, SE-60174 Norrk\"oping, Sweden}

\author{H. Ahmed}\affiliation{Centre for Plasma Physics (CPP), Queen's University Belfast, BT7 1NN, UK}

\author{D. Doria}\affiliation{Centre for Plasma Physics (CPP), Queen's University Belfast, BT7 1NN, UK}

\author{G. Sarri}\affiliation{Centre for Plasma Physics (CPP), Queen's University Belfast, BT7 1NN, UK}

\author{A. Ynnerman}
\affiliation{Department of Science and Technology, Link\"oping University, SE-60174 Norrk\"oping, Sweden}

\author{M. Borghesi}\affiliation{Centre for Plasma Physics (CPP), Queen's University Belfast, BT7 1NN, UK}

\date{\today}

\begin{abstract}
A pair of curved shocks in a collisionless plasma is examined with a two-dimensional particle-in-cell (PIC) simulation. The shocks are created by the collision of two electron-ion clouds at a speed that exceeds everywhere the threshold speed for shock formation. A variation of the collision speed along the initially planar collision boundary, which is comparable to the ion acoustic speed, yields a curvature of the shock that increases with time. The spatially varying Mach number of the shocks results in a variation of the downstream density in the direction along the shock boundary. This variation is eventually equilibrated by the thermal diffusion of ions. The pair of shocks is stable for tens of inverse ion plasma frequencies. The angle between the mean flow velocity vector of the inflowing upstream plasma and the shock's electrostatic field increases steadily during this time. The disalignment of both vectors gives rise to a rotational electron flow, which yields the growth of magnetic field patches that are coherent over tens of electron skin depths.
\end{abstract}

\maketitle

\section{Introduction}

The collision of an ionized blast shell with an ambient plasma triggers the formation of shocks if the collision speed exceeds a threshold value. The critical speed depends on the plasma wave mode that is mediating the shock and on the importance of Coulomb collisions between particles. If the mean frequency, with which the plasma particles collide, is well below all resonance frequencies of the plasma, then the effects of binary collisions are negligible and the shocks are mediated by electrostatic and electromagnetic fields. 

The plasma processes that sustain a collisionless shock and the structure of the associated electromagnetic fields vary strongly between the different plasma regimes. Solar system shocks, like the Earth's bow shock \cite{Bale05,Burgess05}, are immersed in a plasma that is carrying a relatively strong magnetic field and they connect two plasmas that collide at a non-relativistic speed. Such shocks are usually mediated by magnetosonic waves. The collision speed between a supernova blast shell and the ISM at a late evolution phase is similar to the collision speed between the solar wind and the Earth's bow shock. The amplitude of the magnetic field in the ISM, into which a supernova remnant (SNR) shock expands, is weaker by an order of magnitude than that in the solar wind plasma at the Earth's orbit. Magnetosonic waves, which have a low field amplitude, may not be able to sustain permanently the shock because other instabilities develop faster and on a smaller spatial scale. Simulations have shown that drift instabilities and electrostatic turbulence can in some cases suppress the growth of a magnetosonic wave \cite{Dieckmann14}. As we go to higher flow speeds, the plasma shocks become magnetized by filamentation instabilities \cite{Kazimura98,Frederiksen04,Spitkovsky08,Stockem14a}. 

We consider here shocks, which develop in an initially unmagnetized and collisionless plasma. Such shocks are frequently observed in the laboratory \cite{Koopman67,Dean71,Bell88,Romagnani08,Gregori12,Ahmed13} and they might be representative for SNR shocks in their late evolution phase. An electrostatic shock in its most basic form is characterized by a potential difference, which is sustained self-consistently by the plasma. The shock connects the downstream region and an upstream region ahead of the shock. Electrons stream from the denser downstream region into the upstream region and create a charge imbalance between both regions. The denser downstream plasma goes on a positive potential relative to the upstream plasma. 

The upstream plasma streams towards the shock at a speed, which exceeds the ion acoustic speed. The upstream ions are slowed down and compressed by the potential jump as they cross the shock. The potential jump reflects some of the incoming upstream ions, which then move back upstream. The remainder of the incoming ions enters the downstream region, which expands due to the accumulation of the inflowing ions. The shock is thus not stationary in the downstream frame of reference and moves upstream. 

An electrostatic shock is an ion phase space structure, which consists of inflowing upstream ions, reflected ions and ions that overcame the positive potential and accumulated downstream of the shock. The electrostatic field, which mediates the shock, will also accelerate some of the downstream ions into the upstream direction and act as a double layer. Many shocks in unmagnetized plasma are a combination of a double layer and of an electrostatic shock. Such hybrid structures \cite{Hershkowitz81} are characterized by a unipolar electric field. In what follows we shall refer to these hybrid structures as plasma shocks to distinguish them from pure electrostatic shocks.

The incoming upstream ions, which have been reflected by the electrostatic shock, and the downstream ions, which have been accelerated upstream by the double layer, form a beam that outruns the plasma shock. We shall refer to this beam as the shock-reflected ion beam. The number density of this beam can be a significant fraction of that of the incoming upstream ions, which implies that the ion number density ahead of the plasma shock is well above that in the far upstream region. In what follows, we refer to this region as the foreshock. 

The counterstreaming nonrelativistic and unmagnetized ion beams drive the electrostatic ion acoustic instability in the foreshock \cite{Karimabadi91,Kato10}. The speed of the shock-reflected ions in the upstream frame of reference exceeds the ion acoustic speed. The ion acoustic instability can, however, only be destabilized if the beam velocity component along the wave vector is subsonic. The wave vectors of the unstable waves can thus not be parallel to the plasma flow velocity vector. Oblique electrostatic waves grow and the obliquity angle is such that the beam velocity component along the wave vector is comparable to the ion acoustic speed \cite{Forslund70}. Obliquely propagating ion acoustic waves grow in the foreshock region and modulate the incoming upstream ions. The plasma shock is either transformed into a shock with a broad transition layer \cite{Dieckmann14} or it is destroyed by the inflowing turbulent plasma \cite{Kato10}. 

Previous simulation studies have addressed the evolution of (quasi-)planar plasma shocks. The planarity has been enforced in the PIC simulations by resolving only one spatial direction or by choosing initial conditions that are uniform along one direction. However, in particular the plasma shocks in laboratory experiments are often nonplanar. This motivates our study of the formation and evolution of curved shocks with a particle-in-cell (PIC) simulation. 

The shock curvature is introduced in our simulation through the following setup. The two electron-ion clouds collide at a boundary, which is orthogonal to the spatially uniform collision direction at the simulation's start. The mean speed of the plasma along the collision direction varies as a function of the orthogonal direction and this velocity shear gives rise to a shock front that becomes increasingly curved in time. The amplitude of the velocity change is comparable to the ion acoustic speed. 

We find that the shock formation and its stability are not affected by this large velocity shear. The life-time of the plasma shocks is of the order of tens of inverse ion plasma frequencies. The shock transition layer is transformed after this time by the onset of ion acoustic turbulence in the foreshock. The transition from a sharp electron skin depth-scale structure into a broad transition layer is also observed for planar shocks. The key difference between the structure of the curved shock and a planar shock is tied to the disalignment of the electric field with the flow velocity vector of the incoming upstream plasma. The disalignment gives rise to a rotational component of the electron flow, which yields the growth of magnetic field patches. These patches are coherent over tens of electron skin depths and their size is limited by the simulation box dimension. Their amplitude yields a ratio of the electron plasma frequency to the cyclotron frequency of about 100. 

Our paper is structured as follows. Section 2 discusses the PIC code and the initial conditions we use. Section 3 examines the formation and the evolution of the pair of electrostatic shocks. Section 4 summarizes our findings.

\section{Initial conditions and the PIC method}

Particle-in-cell (PIC) simulation codes \cite{Dawson83} solve the Vlasov-Maxwell system of equations via the method of characteristics \cite{Dupree63}. The electromagnetic fields are evolved in time via Amp\`{e}re's law and Faraday's law.
\begin{equation}
\mu_0 \epsilon_0 \frac{\partial \mathbf{E}}{\partial t} = \nabla \times \mathbf{B} - \mu_0 \mathbf{J},
\label{Ampere}
\end{equation}
\begin{equation}
\frac{\partial \mathbf{B}}{\partial t} = - \nabla \times \mathbf{E}.
\label{Faraday}
\end{equation}
Most codes fullfill the equations $\nabla \cdot \mathbf{E} = \rho / \epsilon_0$ and $\nabla \cdot \mathbf{B} = 0$ either as constraints or via correction steps. The plasma is approximated by an ensemble of computational particles (CPs), which correspond to volume elements of the phase space density distribution. The charge-to-mass ratio of the CPs equals that of the plasma particles they represent. Their momentum and position are updated with the relativistic Lorentz force equation
\begin{equation}
\frac{d\mathbf{p}_j}{dt} = q_i \left ( \mathbf{E}(\mathbf{x}_j) + \mathbf{v}_j \times \mathbf{B}(\mathbf{x}_j) \right ).
\label{LorentzForce}
\end{equation}
The CP with the index $j$ of the species $i$ with the position $\mathbf{x}_j$ and the relativistic momentum $\mathbf{p}_j = m_i \Gamma_j \mathbf{v}_j$ has the charge $q_i$ and the mass $m_i$. Its position is updated by $d\mathbf{x}_j / dt = \mathbf{v}_j$. The CPs and the electromagnetic fields $\mathbf{E}(\mathbf{x},t)$ and $\mathbf{B}(\mathbf{x},t)$ are connected as follows. The fields are interpolated to the particle position and update its momentum in time. The current density contribution of each CP is interpolated to the grid. The summation over all current density contributions yields the macroscopic current density $\mathbf{J}(\mathbf{x},t)$, which updates the electromagnetic fields via Amp\`{e}re's law. 

A pair of shocks is generated in the simulation by the collision of two spatially uniform plasma clouds of equal density. Each cloud consists of electrons with the charge $-e$ ($e$: elementary charge), the mass $m_e$, the number density $n_0$ and the temperature $T_e = 1$ keV. We model Deuterium ions with the number density $n_0$, the mass $m_D$ and the temperature $T_D = 200$ eV. We motivate our choice for the initial conditions as follows. 

The collision of two plasma clouds in laboratory- or astrophysical settings usually involves two plasmas with densities that differ by orders of magnitude. The blast shell, which is ejected during a supernova, is composed of stellar material. Its density is thus initially much higher than that of the ambient plasma, which is the stellar wind the star emanated prior to the supernova. A laser-generated blast shell is also much denser than the ambient medium, which is the residual gas that has been ionized by secondary x-ray radiation from the target. 

The blast shells are practically unaffected by the ambient medium during their initial expansion phase and they expand in the form of rarefaction waves. The front of the blast shell becomes faster and thinner in time. Once the expansion speed of the blast shell becomes supersonic and once the ram pressure of the expanding blast shell becomes comparable to the thermal pressure of the ambient medium, the front can be confined and shocks form. Experimental observations \cite{Ahmed13} and PIC simulations \cite{Sarri11} suggest that shocks form when the densities of the colliding clouds become comparable. Selecting equal densities for both colliding plasma clouds should thus be a valid initial condition.

The electron and ion temperatures we use are typical for plasmas, which are created when an ultrashort laser pulse ablates a solid target and if time scales are considered, which are short compared to the time it takes to establish a thermal equilibrium via collisions between electrons and ions. The temperature of the electrons in the ambient plasma, which has been ionized by secondary x-ray emissions from the laser-ablated target, is conparable to 1 keV. The temperatures of the ions of the ambient medium and of the blast shell are usually well below that of the electrons \cite{Eidmann00}.

Deuterium ions have the same charge-to-mass ratio as the fully ionized light atoms, which we usually encounter in laser-plasma experiments, and they have the largest thermal velocity spread for a given temperature. Hence they will provide the strongest ion Landau damping. If we observe the growth of ion acoustic waves for counter-streaming beams of Deuterium, then the same will be true for equally hot plasmas, which consist of heavier ions with the same charge-to-mass ratio as Deuterium. 

The plasma frequencies of the electrons and of the ions are $\omega_{p,e}={(n_0 e^2/\epsilon_0 m_e)}^{1/2}$ and $\omega_{p,i}={(m_e/m_D)}^{1/2} \omega_{p,e}$, respectively. The ion acoustic speed in this plasma is $c_s = {(\gamma_c k_B (T_e + T_D)/m_D)}^{1/2}$ assuming that the adiabatic constant $\gamma_c = 5/3$ is the same for both species. The ion acoustic speed is $c_s=3.1 \cdot 10^5$ m/s. 

The simulation domain has the dimensions $L_x \times L_y = 176 \lambda_s \times 26.2 \lambda_s$ where $\lambda_s = c / \omega_{p,e}$ is the electron skin depth. The boundary conditions are periodic along $y$ and open along $x$. Choosing periodic boundaries along $y$ implies that our simulation evolves in time a periodic chain of blast shells rather than a solitary one. The simulation domain is split up into two equal parts along $x$. The blast shell occupies the interval $-L_x / 2 \le x \le 0$ and $-L_y/2 < y \le L_y/2$. The second cloud, which we refer to as the ambient plasma, occupies the interval $0 < x < L_x / 2$ and $-L_y/2 < y \le L_y/2$.

The initial mean speeds of the electrons and ions of the ambient plasma vanish. The mean speed of the blast shell's electrons equals that of the ions and is denoted here as $\mathbf{v}=(\bar{v}_x,0,0)$. The value of $\bar{v}_x$ varies piecewise linearly along the y-axis. The largest value of $\bar{v}_x = 1.15 \times 10^6$ m/s or $3.7c_s$ is reached at the position $y=0$. The speed decreases linearly in both y-directions. It reaches its minimum of $8.7\times 10^5$ m/s or $2.8c_s$ at the boundaries at $y=-L_y/2$ and $y=L_y/2$.

The simulation box is resolved by 4000 grid cells along the x-direction and by 600 cells along the y-direction. The quadratic side length of each cell is $\Delta_x = 0.044\lambda_s$. Each plasma species is resolved by 200 CPs per cell. We evolve the system for $T_{sim} \omega_{p,i} = 153$ using $3.2 \times 10^5$ time steps $\Delta_t$. In what follows, we normalize time to $1/\omega_{p,i}$, space to $\lambda_s$ and speed to the electron thermal speed $v_{th,e}={(k_B T_e / m_e)}^{1/2}$. The electric field is normalized to $m_e \omega_{p,e} c /e$ and the magnetic one to $m_e \omega_{p,e}/e$. 

\section{The simulation results}

In what follows, we shall present and discuss the spatial distributions of the ion density, of the amplitude of the flow-aligned electric field component $E_x (x,y)$ and the out-of-plane magnetic field distribution $B_z(x,y)$ at the times $t$ = 6, 13.8, 19.4, 33.8, 70 and 153. 

\begin{figure*}
\includegraphics[width=7in]{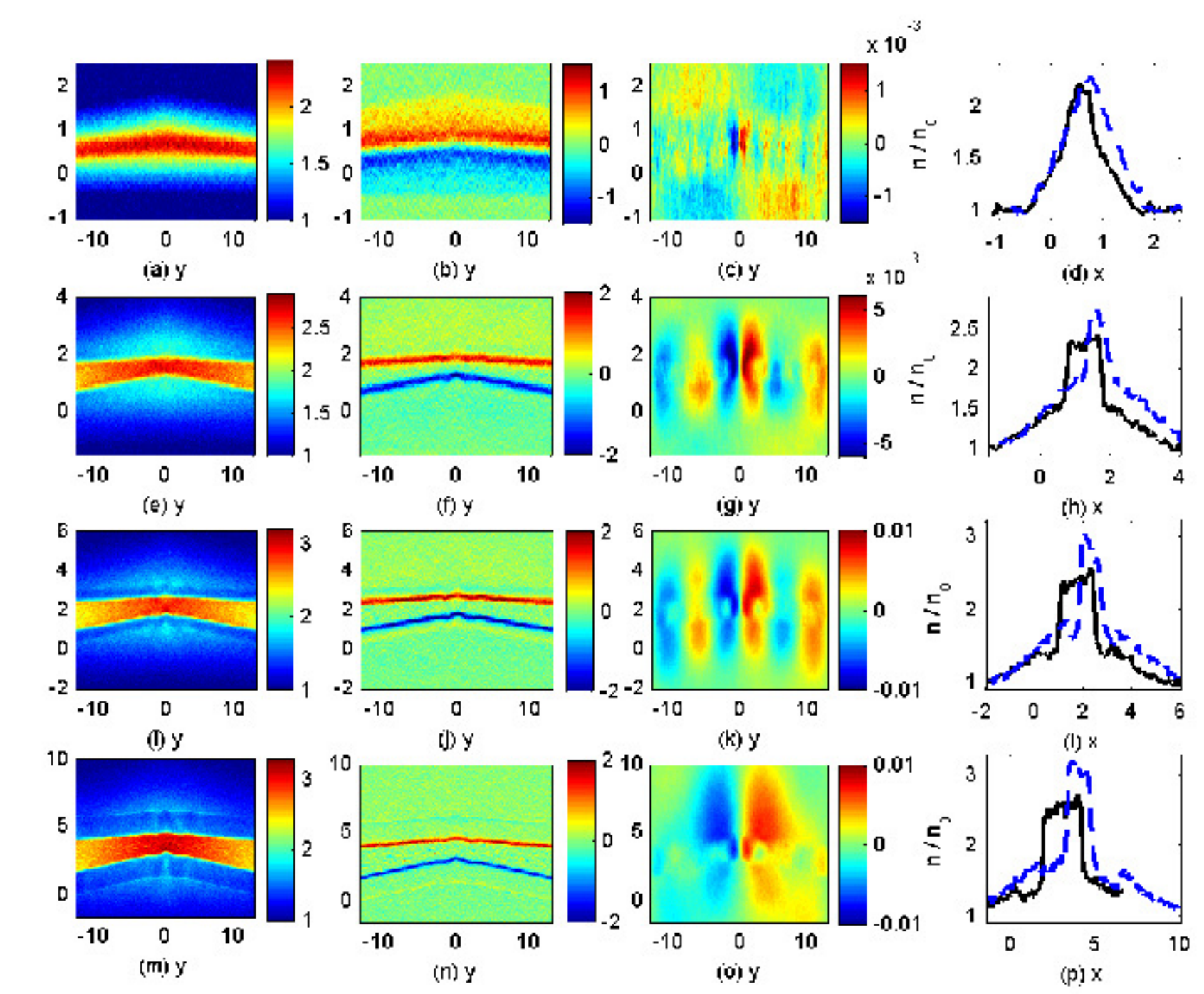}
\caption{The ion density distributions expressed in units of $n_0$ are shown in the first column (from left to right). The distributions of $E_x(x,y)$ are shown in the second column. The electric amplitudes have been multiplied by a factor 100. The third column shows the distributions of $B_z(x,y)$ and the right column shows slices of the ion density distributions along $y=0$ (dashed blue) and along $y=13.1$ (black). The upper row corresponds to the time $t=6$, the second row to $t=13.8$, the third one to $t=19.4$ and the bottom row to $t=33.8$. \label{FourTimes}}
\end{figure*}

Figure \ref{FourTimes} shows the ion density, the electric $E_x$ component and the magnetic $B_z$ component close to the initial contact boundary $x=0$ at several times.
A band with an increased ion density is visible in Fig. \ref{FourTimes}(a). The ions of both clouds interpenetrate and their cumulative density exceeds $n_0$. The peak density of the ions is reached at $x \approx 0.7$ and it exceeds $2n_0$ for all values of $y$. This ion cloud overlap layer is broadest at $y\approx 0$. At this time the ions of both clouds move independently and the width of the layer is proportional to the speed at which the clouds collide. The ion density is not constant close to its maximum value in Fig. \ref{FourTimes}(d). Hence a downstream region, which is characterized by a spatially uniform plasma distribution along $x$ that separates the forward and reverse shocks, has not formed at this time. 

A strong bipolar electric field pulse is visible in Fig. \ref{FourTimes}(b). It is sustained by the space charge, which results from the electrons that escaped from the ion cloud overlap layer. The polarity of the electric field is such that the ion cloud overlap layer is on a positive potential relative to the ambient- and blast shell plasmas. The slow-down of the inflowing ions by this potential is responsible for the increase of the ion density beyond $2n_0$. 

Figure \ref{FourTimes}(c) reveals magnetic oscillations within the ion cloud overlap layer. The strongest oscillations are located at $y \approx 0$ and $x\approx 0.7$, where the plasma collides at the highest speed. The growth of $B_z$ within the ion cloud overlap layer can be attributed partially to the instability observed by \cite{Stockem14b} for shocks and by \cite{Thaury10} and \cite{Quinn12} for rarefaction waves. The electrons are accelerated along the electric field of the ion cloud overlap layer. The resulting directional anisotropy in the velocity distribution triggers the Weibel instability \cite{Weibel59,Stockem09}. 

\begin{figure}
\includegraphics[width=\columnwidth]{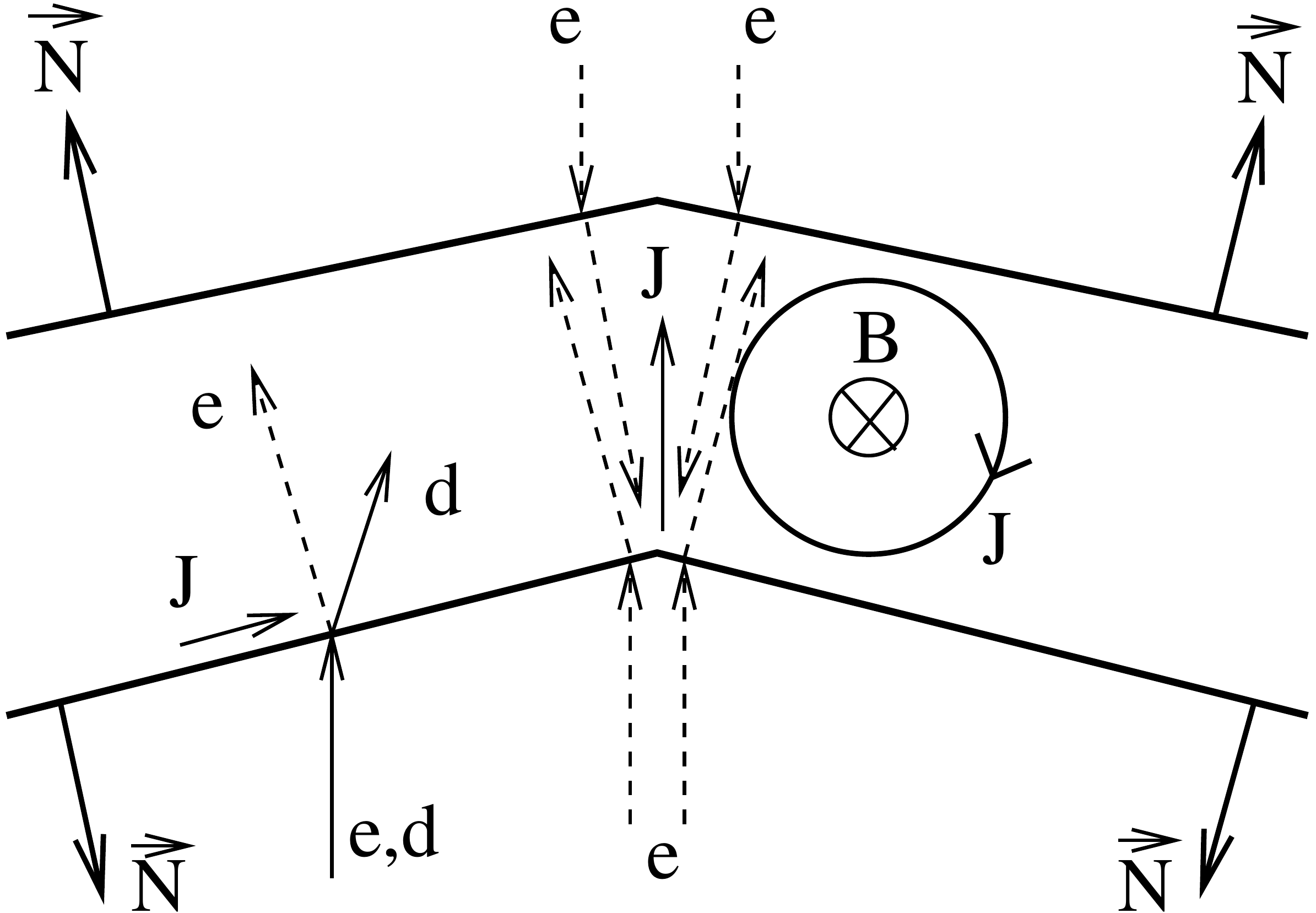}
\caption{The field and flow diagram within the overlap layer of both plasma clouds. The normals of the boundary between the overlap layer and the upstream plasma are denoted by $\mathbf{N}$ and are parallel to the vector of the ambipolar electric field. Electrons are denoted by $e$ and their velocity vectors are the dashed arrows. Deuterium ions are denoted by $d$ and their velocity vectors are solid. Net currents are denoted by solid vectors and the symbol $J$. The circulating current gives rise to a magnetic field (Symbol B).}\label{Sketch}
\end{figure}

The Weibel instability in the form discussed in Ref. \cite{Stockem09} can, however, not explain why the magnetic field oscillations peak at $y\approx 0$ and $x\approx 0.5$. We attribute this to the geometric effect that is outlined in Fig. \ref{Sketch}.

We consider first the boundaries between the overlap layer and the upstream plasma that have a constant slope. Electrons that stream into the overlap layer are accelerated by the ambipolar electric field that ensheaths the overlap layer. The field is aligned with the boundary normal. The electrons are accelerated and the ions are decelerated when they enter the overlap layer, while their lateral velocity components remain unchanged. The velocity vectors of the inflowing electrons and ions are thus rotated into opposite directions. Their current contributions do no longer cancel each other out and a net current develops within the overlap layer.

An even stronger net current develops close to the cusps due to the changing direction of the normal. Electrons that cross the overlap layer at a concave cusp are scattered. Electrons that cross it at a convex cusp are focused. The electron current is no longer balanced along the vertical direction in the center of Fig. \ref{Sketch}. A net current flows from the concave to the convex cusp of the overlap layer and a magnetic field will grow, which is at least initially confined to the overlap layer. The growing spatially localized magnetic field will induce a ring current within the overlap layer. The direction of the magnetic field in Fig. \ref{Sketch} matches that in the simulation if we take into account that the z-axis points into the plot's plane in Fig. \ref{FourTimes}(c).

Magnetic field oscillations are present on both sides of the ion cloud overlap layer in Fig. \ref{FourTimes}(c). Their wavelength is $2\pi/L_y$ along y. This modulation extends up to the boundary at $-L_x/2$ and it does not oscillate along the x-direction (not shown). Such an oscillation can not be driven by an electron current that emanates from the overlap layer. Electrons that move at the thermal speed can only traverse the distance $\approx 15 \lambda_s$ during $t=6$. Hence the magnetic field must be driven by spatial and temporal variations of the electric field and propagate in the light mode. Faraday's law gives $\frac{\partial}{\partial t} B_z = \frac{\partial}{\partial y} E_x - \frac{\partial}{\partial x} E_y$. The observation of a magnetic field $B_z \neq 0$ indicates that the condition for a two-dimensional electrostatic potential $\partial E_x / \partial y = \partial E_y / \partial x$ is not fulfilled by the bipolar structure in Fig. \ref{FourTimes}(b). The noise and the Weibel instability, which triggers the growth of patchy magnetic fields in the ion cloud overlap layer of the ions, introduce a weak magnetic component of the shock to start with. The magnetic fields, which grow upstream of both shocks, remain weak and we shall not discuss them further. 

Figures \ref{FourTimes}(e-h) show the ion- and electromagnetic field distributions sampled at $t=13.8$. The ion cloud overlap layer in Fig. \ref{FourTimes}(e) is now broader and less dense at the boundaries $y=\pm L_y/2$ than in the center, which is demonstrated quantitatively by Fig. \ref{FourTimes}(h). The ion density in Fig. \ref{FourTimes}(h) downstream of the slow shocks evidences a flat density profile in the interval $0.7 < x < 1.8$, which is typical for a downstream region. 

The larger plasma compression by fast shocks compared to that of slow shocks implies that the density ratio between the downstream plasma and the upstream plasma is larger for the shocks close to $y\approx 0$ in Fig. \ref{FourTimes}(e). Fast shocks like the ones at $y\approx 0$ reflect most of the incoming ions and only a minor fraction enters the downstream region \cite{Forslund71,Dieckmann13}. Consequently, the downstream region of the fast shocks expands slowly. More of the incoming upstream ions can traverse the slow shocks close to $y=\pm L_y/2$ and the downstream region behind them accumulates more ions. It expands faster. The ion density in Fig. \ref{FourTimes}(h) is decreasing rapidly to a value $\approx 1.5n_0$ at $y=\pm L_y/2$ as $x$ is decreased below 0.7 or increased above 1.8. The density converges to $\approx n_0$ at the boundaries of the displayed interval. The density enhancements between $-1<x<0.5$ and $2<x<4$ are caused by the shock-reflection of ions and by ions that propagated through the ion cloud overlap layer before both shocks formed.

The electric field outlines the location of the forward and reverse shocks in Fig. \ref{FourTimes}(f). The unipolar electric field pulses, which mediate the shocks, are closest at $y\approx 0$. Their separation increases along $x$ as we go to the boundaries at $y= \pm L_y/2$. The thickness of the unipolar electric field peaks along $x$ is less than in Fig. \ref{FourTimes}(b). This change of the thickness of the pulse evidences the transformation of the ion cloud overlap layer into a downstream region that is enwrapped by forward and reverse shocks. 

The magnetic field amplitude modulus in Fig. \ref{FourTimes}(g) has quadrupled at $y \approx 0$ compared to that at the earlier time. Additional field structures have emerged close to the boundaries at $y=\pm L_y/2$ and at $y=\pm L_y / 4$. The magnetic field is strongest inside of the downstream region. It is, however, not confined by it like in the simulation in Ref. \cite{Stockem14b}. The strongest magnetic fields are observed in the intervals along $y$, where the cloud collision speed has extrema and where the overlap layer has cusps. The normalization of the fields implies that the peak amplitude of the magnetic field yields a ratio of the electron cyclotron frequency to the electron plasma frequency of about $6 \times 10^{-3}$.

The ion distribution has changed qualitatively at the time $t=19.4$, which is evidenced by the Figs. \ref{FourTimes}(i-l). The ion density in the downstream region is comparable to that at the earlier time $t=13.8$, but the ion density is now also a function of $x$. The density of the downstream ions in Fig. \ref{FourTimes}(i) and in Fig. \ref{FourTimes}(l) is highest at the concave shocks and lowest at the convex shocks. The ion density at $y=0$ and $x\approx 2$ is about $3n_0$ at the concave reverse shock and it decreases to a value of $2.7n_0$ at $x\approx 2.5$. The electric field in Fig. \ref{FourTimes}(j) reacts to it because a larger change of the ion density across the concave shock yields a stronger ambipolar electric field. The electric field modulus at the concave shock at $y=0$ and $x\approx 1.8$ exceeds that of the convex shock at $x\approx 2.5$. The magnetic $B_z$ component in Fig. \ref{FourTimes}(k) component shows a cellular structure and peak amplitudes of $\approx 8\times 10^{-3}$ are reached close to $y\approx 0$. 

Figures \ref{FourTimes}(m-p) show the ion density distribution and the electromagnetic field distributions at the time $t=33.8$. The ion density distribution in Fig. \ref{FourTimes}(m) and the two density slices shown in Fig. \ref{FourTimes}(p) demonstrate that the density of the downstream plasma immediately behind a shock still depends on whether it is concave or convex. The ion density decreases with increasing $x$ close to $y=0$, while the opposite is true close to $y=\pm L_y/2$. 

The ion density distribution in the foreshock region of each shock has changed from a diffuse distribution in Fig. \ref{FourTimes}(i) to one that shows a pronounced peak that is located about $2\lambda_s$ ahead of each shock. These structures yield a thin band in Fig. \ref{FourTimes}(n) with an amplitude modulus of about $5 \times 10^{-3}$. The electric field of such a structure and the electric field of the nearest shock have the opposite polarization. 

The cellular magnetic field structures from Fig. \ref{FourTimes}(k) have merged to a large magnetic field distribution in Fig. \ref{FourTimes}(o), which extends far into the upstream regions of both shocks. The magnetic field amplitude peaks in the downstream region close to $y\approx 0$. A weaker similar distribution exists close to the boundary at $y=\pm L_y/2$. The shape of the magnetic field structure suggests that the associated current has its source in the kink in the overlap layer at $y=0$. We expect that the net current at this kink is higher than that at $y=L_y/2$, because the plasma flow speed is highest at $y=0$ and because the ion density and, hence, the potential of the overlap layer peak at $y=0$. 

\begin{figure}
\includegraphics[width=\columnwidth]{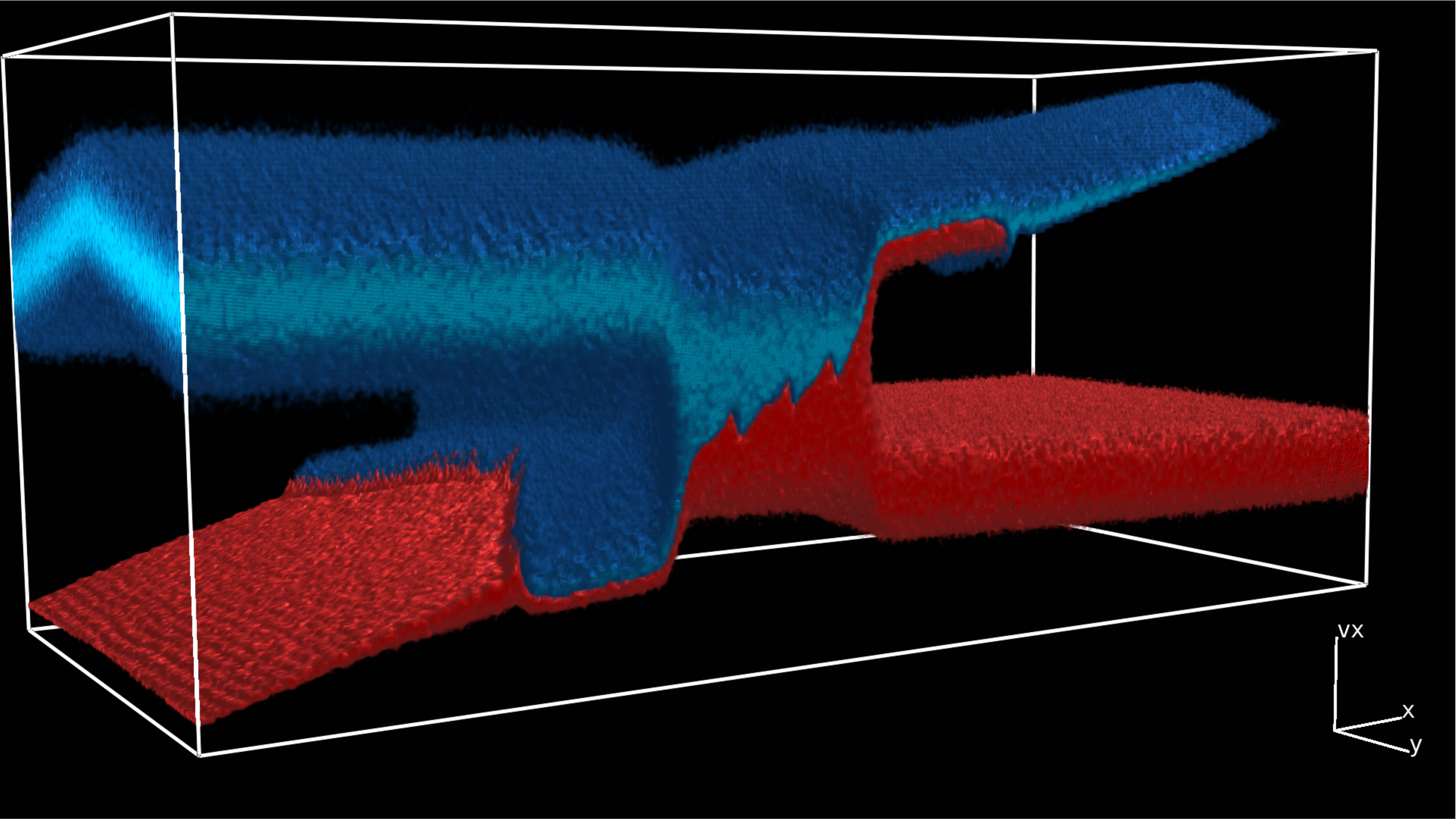}
\caption{The ion phase space density distribution $f_i (x,y,v_x)$ at the time $t=33.8$. The blast shell ions (blue) are located mainly at high positive values of $v_x$ while the ambient ions (red) are located mainly at low values of $v_x$. The x-axis range is [-2.75,11.0], the y-axis range is [-13.1,13.1] and the $v_x$-axis range expressed in units of $c_s$ is [-1.9,5.8].}\label{Rendering175Total}
\end{figure}

The shock structure and the source of the ion density peaks ahead of the main shocks in Fig. \ref{FourTimes}(m) is revealed by the ion phase space density distribution $f_i (x,y,v_x)$, which is displayed in Fig. \ref{Rendering175Total}. 
This distribution shows several distinct features. The incoming blast shell ions are located at low values of $x$ and at high positive speeds $v_x$. The blast shell ions have their largest mean speed at $y\approx 0$ and their speed decreases linearly as we go towards the periodic boundaries in the y-direction. The blast shell propagates to increasing values of $x$. The ambient ions are located at the right (large values of $x$) and their mean speed is zero. The ions of both clouds merge to a structure, which is characterized by a large spread along the $v_x$-axis. This is the downstream ion population and it is bounded by the forward and reverse shocks along both $x$-directions. The velocity change between the blast shell plasma and the downstream region and between the downstream region and the ambient plasma are caused by the ion acceleration by the shock's electric field. 

Each plasma shock gives rise to a shock-reflected ion beam. This beam is composed of the ions that were reflected by the electrostatic shock and of the ions that were accelerated by the double layer as they moved from the downstream region into the upstream region. Let us consider the shock-reflected ion beam, which is located to the left and at low speeds. This beam reveals two distinct regions. The part to the left is uniform along the y-direction and it contains only ions from the ambient plasma. Its mean velocity and its density decreases as we go to decreasing values of $x$. Such a phase space profile is that of a rarefaction wave \cite{Thaury10}. The shock-reflected ion beam close to the shock is no longer spatially uniform along $y$. The boundary between these two regions follows the shock profile and it separates the rarefaction wave, which contains only ions from the ambient plasma, from the shock-reflected ion beam that contains ions from the blast shell plasma and the ambient plasma. The shock-reflected ion beam at large values of $x$ and at positive $v_x$ also shows a clear subdivision into two domains, which are separated by a boundary. The mean velocity of the ions changes across this boundary and the ion acceleration is accomplished by the electric field pulse seen in Fig. \ref{FourTimes}(n) at large $x$.

\begin{figure}
\includegraphics[width=\columnwidth]{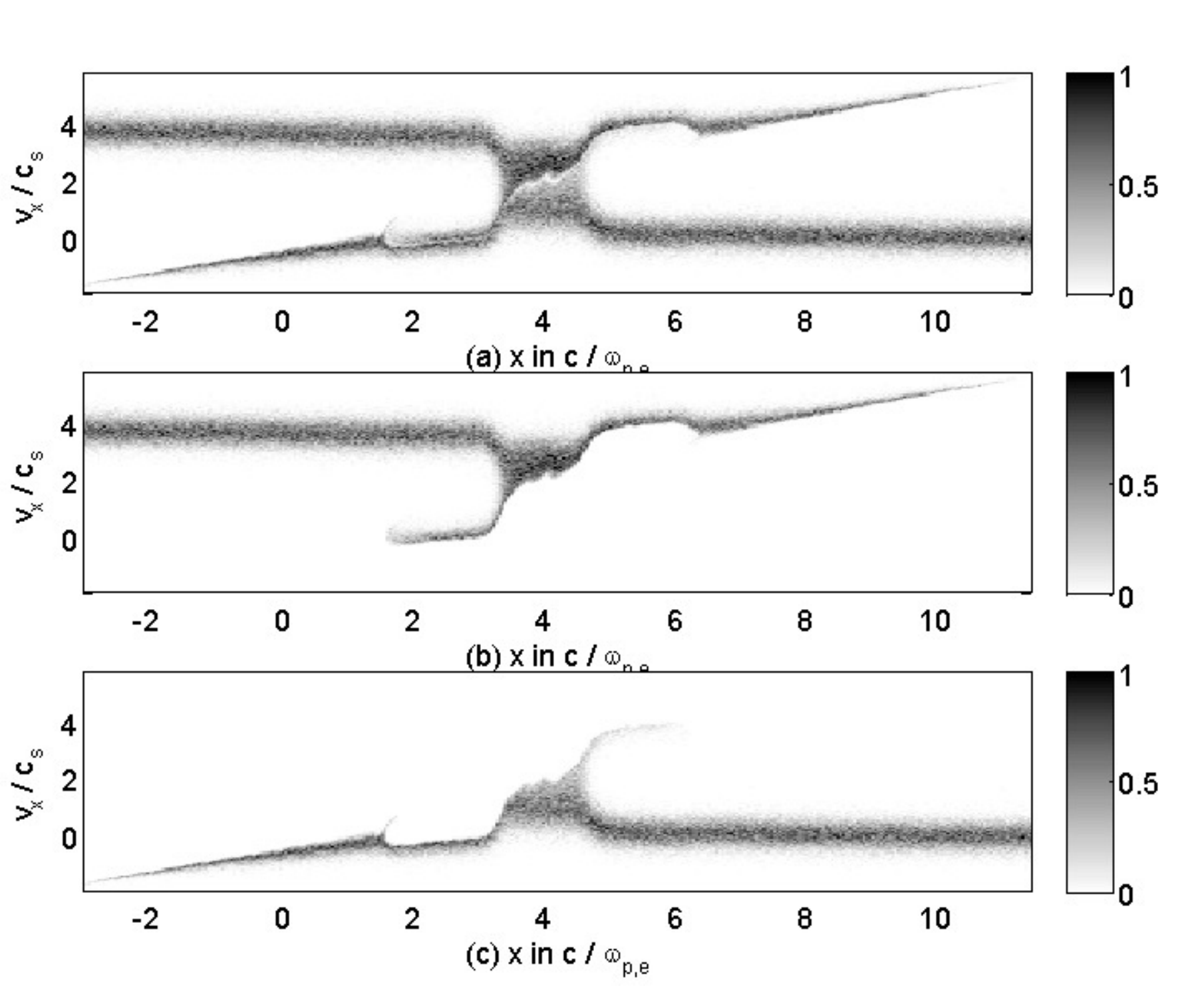}
\caption{The ion phase space density distribution $f_i(x,y,v_x)$ at the time $t=33.8$ and along the slice $y=0$. Positions are normalized to $\lambda_s$ and the velocity axis is normalized to $c_s$. Panel (a) shows the total ion density. Panel (b) shows the blast shell ions and panel (c) the ambient ions.}\label{Plot175PhaseSpaceFast}
\end{figure}

A slice of the ion phase space distribution shown in Fig. \ref{Rendering175Total} along $x$ and for $y=0$ is displayed in Fig. \ref{Plot175PhaseSpaceFast}.
Figure \ref{Plot175PhaseSpaceFast}(a) shows the cumulative ion distribution. The blast shell ions are found in the interval $x < 3$ and $v_x / c_s \approx 4$ and the ambient ions at $x > 5$ and $v_x / c_s \approx 0$. The downstream region is confined to $3.5< x < 4.5$. The blast shell ions in Fig. \ref{Plot175PhaseSpaceFast}(b) are partially reflected at $x\approx 3.5$ and some of them enter the downstream region. This phase space structure is an electrostatic shock. Some of the ions have crossed the downstream region and they have been accelerated by the double layer at $x\approx 4.5$. The beam of accelerated ions has an almost constant speed up to $x\approx 6$, where the beam speed decreases. The beam speed grows linearly and the beam density decreases as we go from $x\approx 6.5$ to $x \approx 10.5$. The latter phase space structure is a rarefaction wave. The sudden decrease of the mean speed of the ion beam at $x\approx 6$ in Fig. \ref{Plot175PhaseSpaceFast}(b) implies that ions at lower $x$ catch up and collide with ions at higher values of $x$. 

\begin{figure}
\includegraphics[width=\columnwidth]{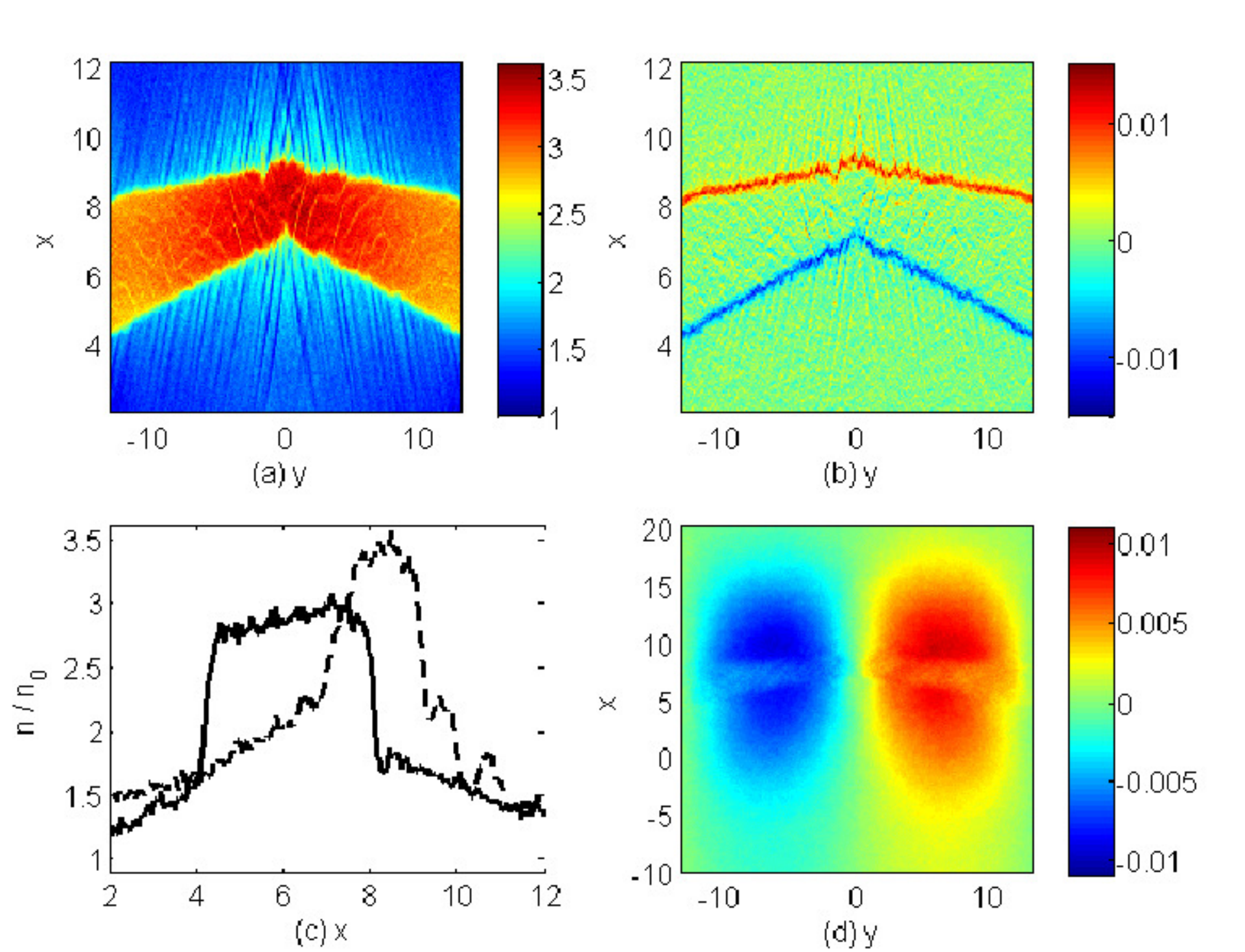}
\caption{The ion density distribution expressed in units of $n_0$ is shown in panel (a). Panel (b) shows the distribution of $E_x(x,y)$. The ion density distributions along two slices $y=0$ (dashed black) and $y=13.1$ (black) are shown in panel (c). Panel (d) shows the distribution of $B_z(x,y)$. The time is $t=70$.\label{Plot366}}
\end{figure}

Figure \ref{Plot366} shows the ion density distribution and the electromagnetic field distributions at the time $t=70$.
The downstream region close to $y\approx 0$ is no longer bounded by smooth shocks. An ion density cusp between the shock and the blast shell has developed at $x\approx 7$ and $y\approx 0$, while the shock is convex on the other side of the downstream region and surrounded by cusps at $y=\pm 2$. The fastest shocks show a more complex distribution than their slower counterparts close to $y\approx \pm L_y/2$. The ion density distribution along $x$ in Fig. \ref{Plot366}(c) of the latter is qualitatively similar to that at previous times; the ion density at the concave shocks exceeds the one at the convex shocks. The amplitude modulus of the electric field peaks, which mediate both shocks, has decreased from a value $\approx 2\times 10 ^{-2}$ at $t=33.8$ to an amplitude modulus of about $1.5 \times 10 ^{-2}$ at $t=70$. The extent of the magnetic field patches in Fig. \ref{Plot366}(d) is now of the order of 10 $\lambda_s$. Any further expansion of the magnetic field patches along y is impeded by the periodic boundary conditions along this direction. 

\begin{figure}
\includegraphics[width=\columnwidth]{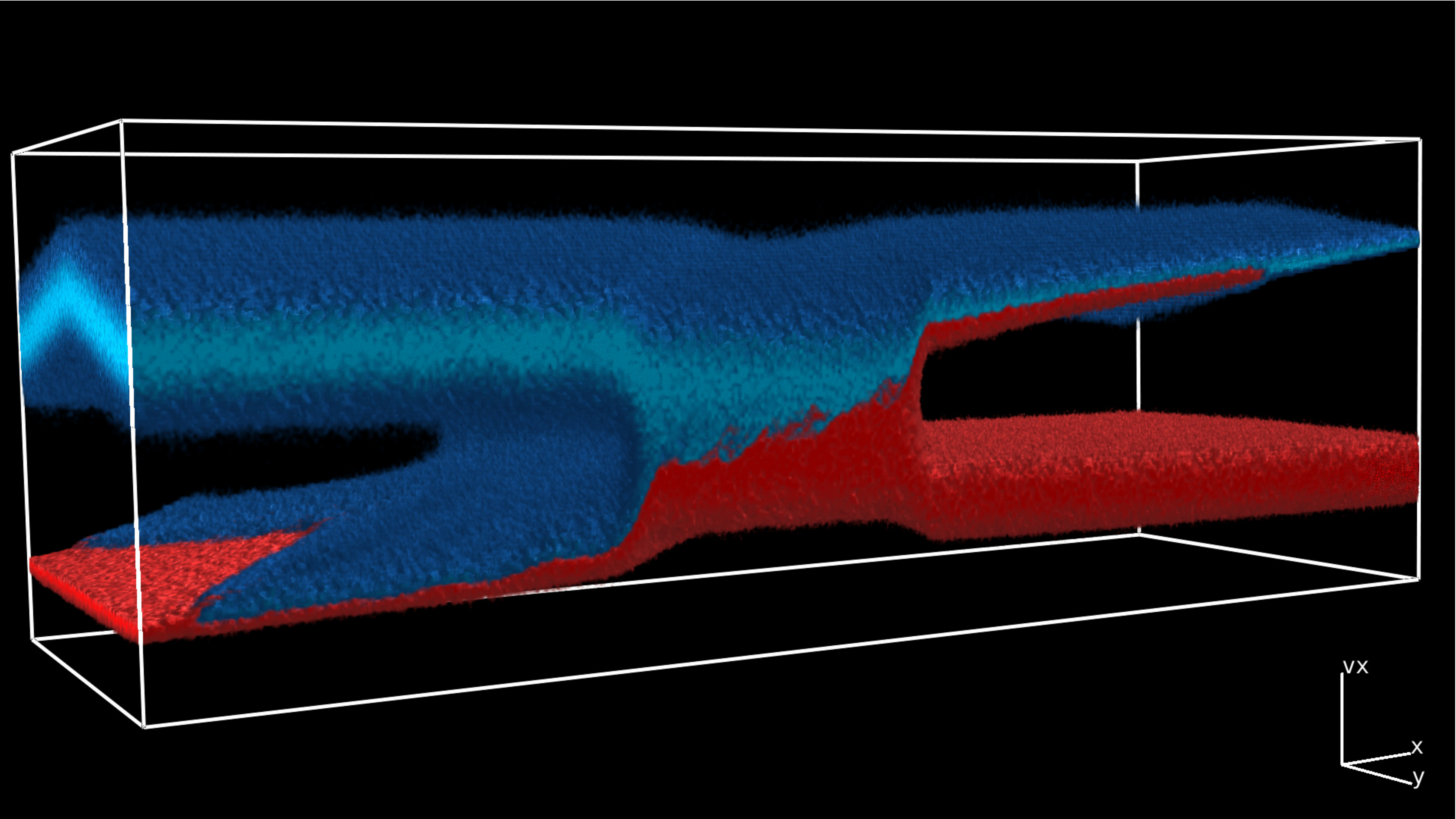}
\caption{The ion phase space density distribution $f_i (x,y,v_x)$ at the time $t=70$. The blast shell ions (blue) are located mainly at high positive values of $v_x$ while the ambient ions (red) are located mainly at low values of $v_x$. The x-axis range is [-5.75,12.0], the y-axis range is [-13.1,13.1] and the $v_x$-axis range expressed in units of $c_s$ is [-1.9,5.8] (multimedia view).}
\label{Rendering365Total}
\end{figure}

The ion phase space density distribution at $t=70$ in Fig. \ref{Rendering365Total} looks qualitatively similar to that at the earlier time $t=33.8$. The ions have started to mix in the downstream region enclosed by both shocks. This mixing is accomplished by the ion phase space vortices, which we can observe close to the interface between the ambient and blast shell ions.
The shock-reflected ions have propagated farther away from the shocks, but there is still a clear subdivision into ions, which were accelerated by the double layers before the shocks formed, and ion beams that were accelerated after the shock formation. The latter consist of ions of the blast shell plasma and of the ambient ions. The shock-reflected ambient ions to the right of the figure have started to overtake some of the blast shell ions.

\begin{figure}
\includegraphics[width=\columnwidth]{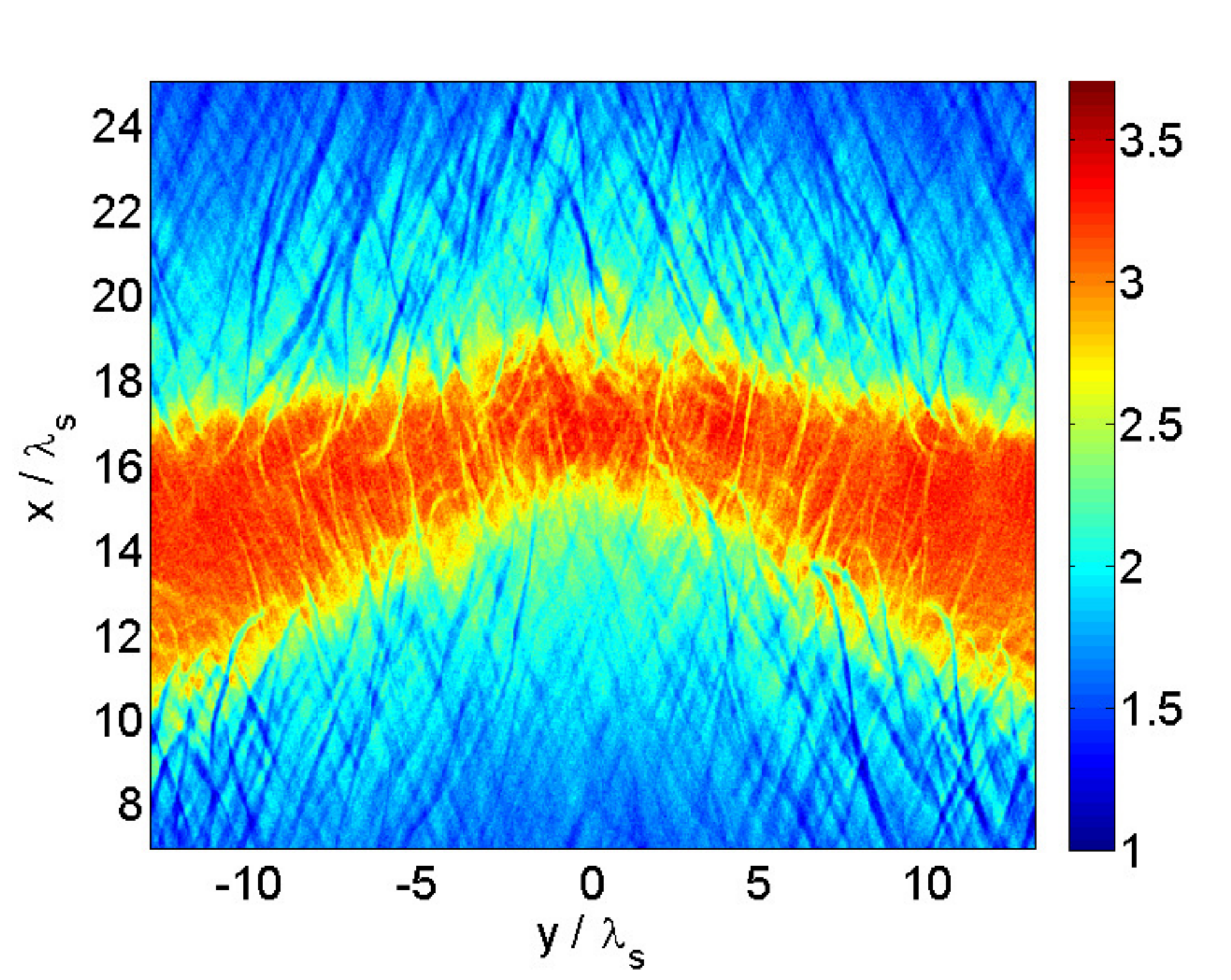}
\caption{The ion density distribution expressed in units of $n_0$ at the time $t=153$ (multimedia view).\label{Plot799a}}
\end{figure}

Figure \ref{Plot799a} shows the ion density distribution at $t=153$.
The downstream region in Fig. \ref{Plot799a} has a density that does no longer vary as a function of $y$. The density in the entire overlap layer is about that observed earlier close to the fastest shocks. Thermal diffusion is one way to equilibrate the ion density in the downstream region. The ion diffusion length can be estimated by multiplying the simulation time $t=153$ with the initial thermal speed $\approx 10^5$ m/s of the ions. Ions moving at this speed can cross the distance $\delta_m \approx L_y / 8$. The shock-heated ions in the downstream region have speeds well in excess of the initial thermal speed and they can cross the distance from the high density region at $y=0$ to the low density region at $y=\pm L_y/2$ during the simulation time. The equilibration of the ion density in the downstream region can thus be accomplished by the thermal diffusion of ions.

\begin{figure}
\includegraphics[width=\columnwidth]{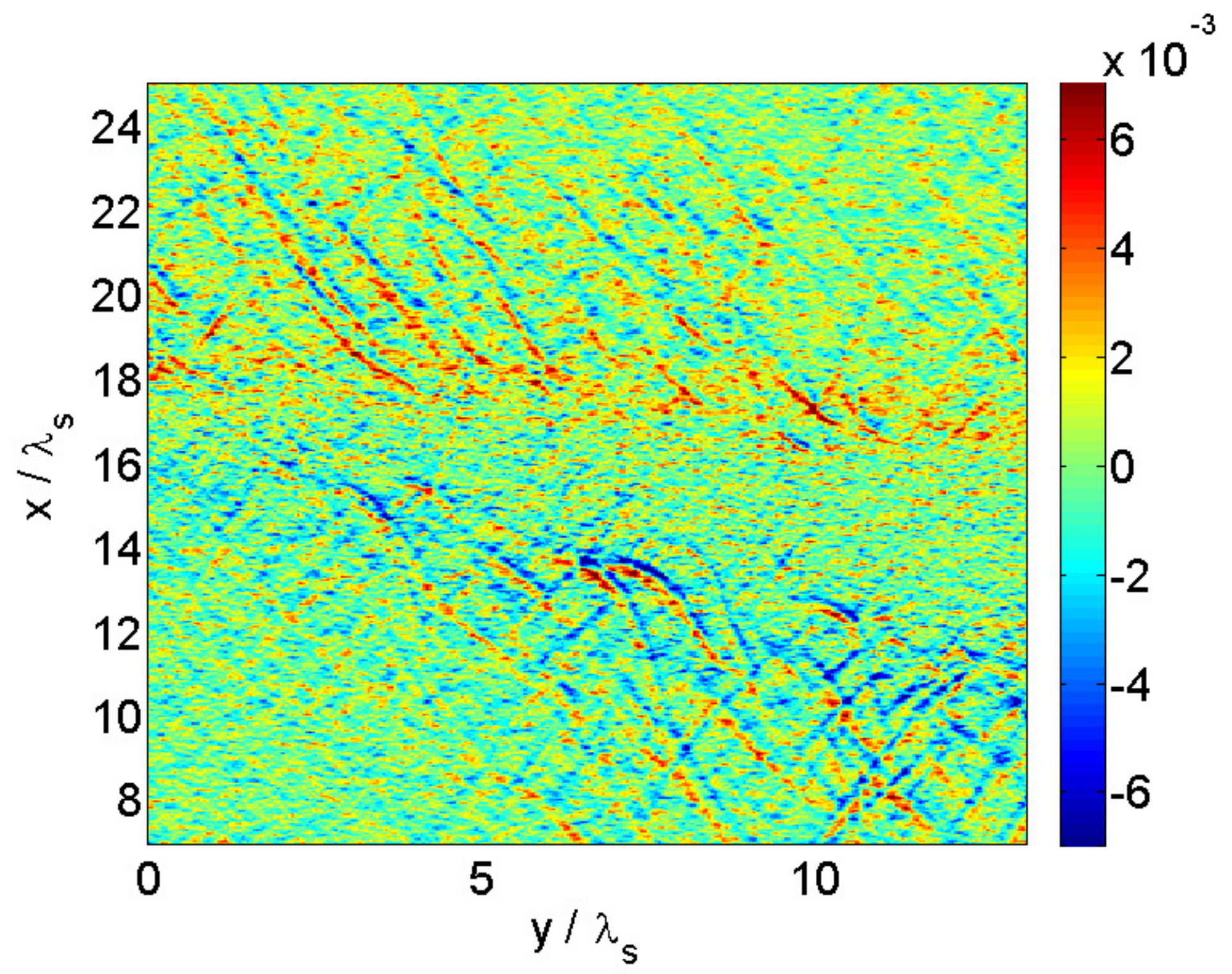}
\caption{The spatial distribution of the electric $E_x$-component at $t=153$ (multimedia view).\label{Plot799b}}
\end{figure}

Figure \ref{Plot799b} shows the distribution of the electric $E_x$ component at $t=153$.
Figure \ref{Plot799b} reveals that the shock transition layer has changed from being a narrow unipolar electric field pulse observed at $t=70$ to a broad layer of electrostatic waves. Such a turbulent layer is driven by ion acoustic waves, which reach electric field amplitudes that are about 50\% of that of the electrostatic shock in Fig. \ref{Plot366}(b). Such a turbulence layer is capable of thermalizing the incoming ions in all directions, since the ions are exposed to a series of strong electric field pulses with an almost random polarization. The broader shock transition layer also results in a lower ion density gradient in Fig. \ref{Plot799a}. A decrease in the ion density gradient brings with it a low amplitude of the ambipolar electric field. That is the reason for why we can no longer see the shock's unipolar electric field even though the potential difference between the downstream region and the upstream region has not changed compared to that at $t=70$.

\begin{figure}
\includegraphics[width=\columnwidth]{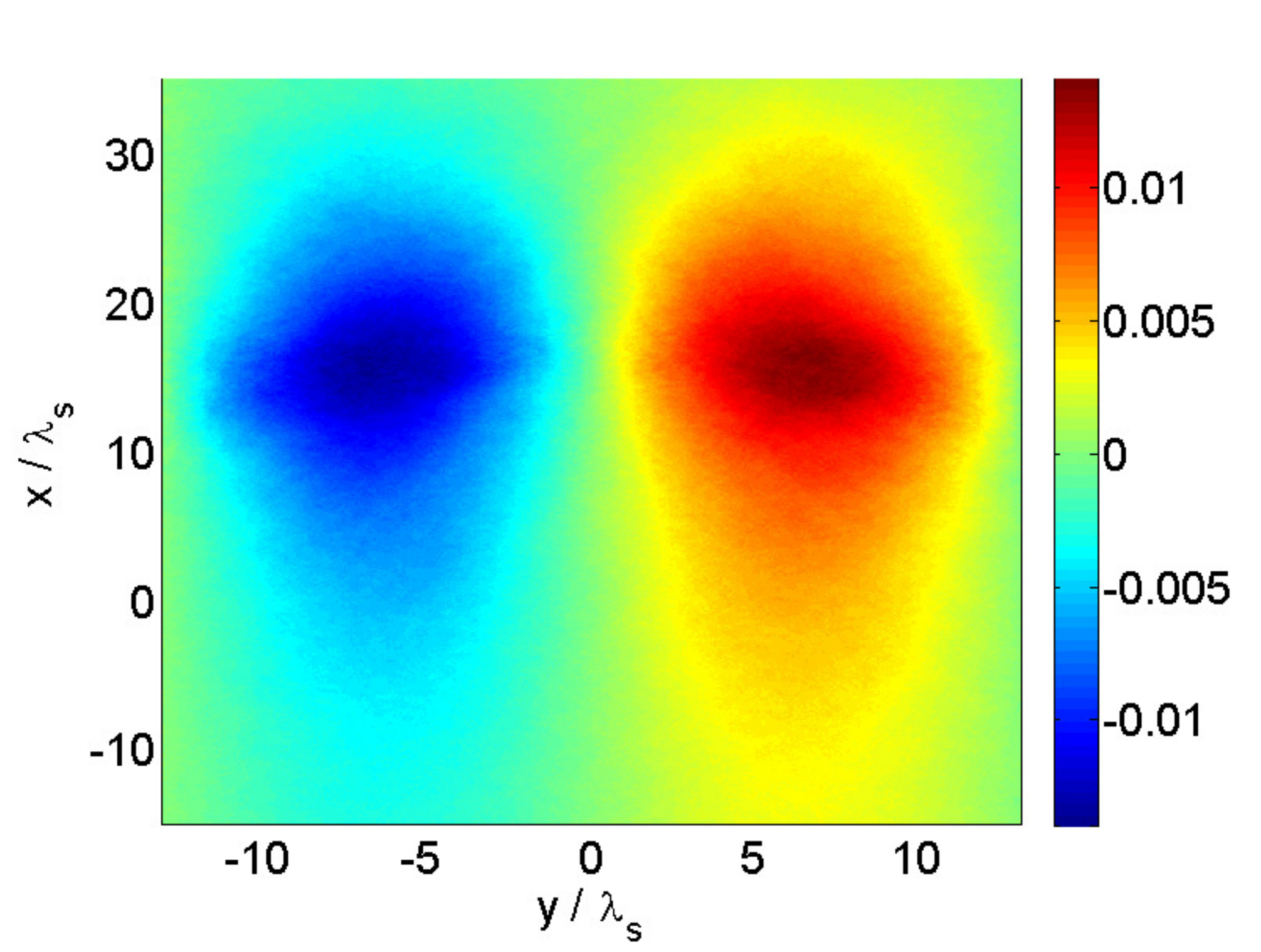}
\caption{The spatial distribution of the magnetic $B_z$-component at $t=153$ (multimedia view).\label{Plot799c}}
\end{figure}

Figure \ref{Plot799c} shows the distribution of the magnetic $B_z$ component at $t=153$.
The magnetic field patches in Fig. \ref{Plot799c} have expanded along $x$ compared to those at $t=70$ and their width along this direction is about $40 \lambda_s$. Their expansion along the y-direction was already limited by the simulation box size at $t=70$ and hence the patches could not expand further along this direction. The coherence scale of the magnetic field patches, their expansion far upstream of the shock and their close correlation with the cusps of the overlap layer exclude the Weibel instability as the cause. The magnetic fields driven by the Weibel instability oscillate in space and their wavelength is comparable to an electron skin depth.

\begin{figure}
\includegraphics[width=\columnwidth]{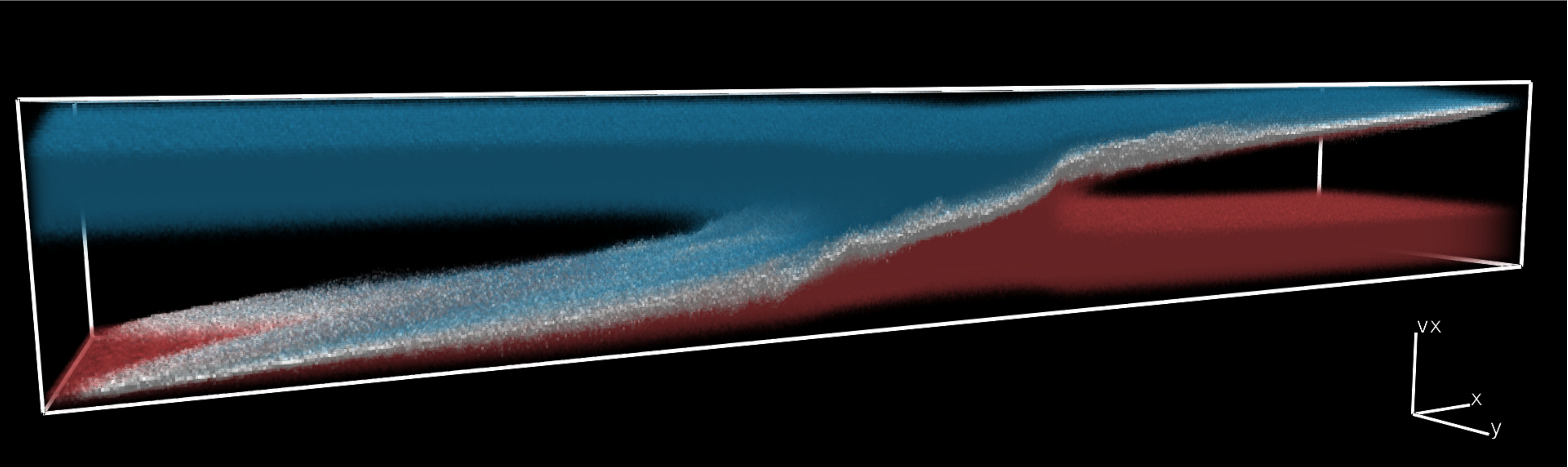}
\caption{The ion phase space density distribution $f_i (x,y,v_x)$ at the time $t=141$. The blast shell ions (blue) are located mainly at high positive values of $v_x$ while the ambient ions (red) are located mainly at low values of $v_x$. The white band shows phase space intervals that are occupied by ions from the blast shell plasma and from the ambient plasma. The x-axis range is [-8.8,22], the y-axis range is [-13.1,13.1] and the $v_x$-axis range expressed in units of $c_s$ is [-1.9,5.8].}
\label{RenderingTotal799}
\end{figure}

The ion phase space density distribution at $t=141$ is shown in Fig. \ref{RenderingTotal799}.
This distribution demonstrates that a downstream region still exists. The shocks, which enclose this region, are more diffuse than at the earlier times. This is a consequence of the ion acoustic turbulence, which is now mediating the shocks. The interaction of the blast shell ions and the ambient ions with the turbulent wave fields mixes both populations in phase space. The phase space interval, in which both ion species have mixed, is indicated with the white band in Fig. \ref{RenderingTotal799}. The white interval corresponds to voxels, in which we find ions from both plasma clouds. 

\section{Summary}

We have examined the formation and evolution of a pair of shocks in an initially unmagnetized plasma. Two plasma clouds, which consisted of electrons and ions, collided at a boundary, which was initially planar. The electrons and ions of both clouds had the same density. The electrons of both clouds had the same temperature. The electron temperature exceeded that of the ions by a factor of 5. The electrons and ions of each cloud had the same mean speed at any position and the plasma was initially free of any net charge and current. The collision speed normal to the initial collision boundary varied as a function of the position along the boundary. The collision speed was highest in the center of the simulation box and decreased linearly with a decreasing distance to the periodic boundary, on which the collision speed reached its minimum value. 

We have obtained the following results. A pair of shocks can form in a plasma with a large velocity shear. These hybrid structures that consist of an electrostatic shock and of a double layer have a lifetime that is comparable to that of planar shocks. The shock normal is initially anti-parallel to the velocity vector of the incoming upstream ions. The incoming upstream ions are slowed down and compressed along the shock normal direction and no particle acceleration takes place in the plane that is orthogonal to the shock normal. The plasma dynamics involves only the position and velocity along the shock normal and the shock dynamics is one-dimensional. Consequently the variation of the shock speed in the downstream frame of reference and the plasma compression with the collision speed resembles that in the parametric study of one-dimensional plasma shocks of \cite{Dieckmann13}.

The shock dynamics becomes two-dimensional after about 70 inverse ion plasma frequencies. The ion density differences in the downstream region are equilibrated by thermal diffusion. The density of the downstream region equilibrates at the previously highest value, which was reached behind the fastest plasma shock. The transition layer of the plasma shock is transformed from a sharp unipolar electric field pulse into a broad layer of electrostatic turbulence. The latter converts the directed flow energy of the upstream ions into heat along all three velocity directions, which leads to a full thermalization of the inflowing upstream ions by the shock.

The formation of the plasma shock and the associated increase of the ion compression yields a sudden increase of the positive potential of the double layer. The ions that cross the double layer towards the upstream direction are accelerated to a higher speed after the shock has formed and they catch up with the ions that crossed the double layer at an earlier time. The magnitude of the velocity change is here sufficient to trigger the formation of a shock, which is located ahead of the main shock and has a life-time of the order of a few inverse ion plasma frequencies. 

The formation of a secondary shock ahead of the primary one has been observed by \cite{Hansen06}. The secondary shock has been attributed the heat wave, which outran the radiative shock in this experiment. We can compare the life-time of this secondary shock to that we have observed in our simulation. The residual gas in Ref. \cite{Hansen06} consisted of a mixture of Xenon and Nitrogen gas with a mass density of $3.6 \times 10^{-5} \textrm{g} \, \textrm{cm}^{-3}$. Let us consider the case study in Ref. \cite{Hansen06}, where the residual gas consists entirely of nitrogen and we furthermore assume that the nitrogen is fully ionized. We obtain an ion plasma frequency $\omega_i \approx 3 \times 10^{12} s^{-1}$. Our simulation would cover a time scale of about 50 ps, which is more than three orders of magnitude shorter than the life-time of the second shock in Ref. \cite{Hansen06}. A lower ionization state of the nitrogen and the presence of neutral nitrogen would reduce the ion plasma frequency and extend the life-time of the secondary shock. A mix of nitrogen with different ionization states may also affect this life-time. It is, however, unlikely that the life-time of the secondary shock could be extended by a factor of 1000. The secondary shock, which has been observed in Ref. \cite{Hansen06}, can thus not be explained in terms of the subshock we have observed here. 

We observed the growth of large magnetic field patches in our simulation. Initially the magnetic field growth was limited to the ion cloud overlap layer. Magnetic fields can be generated via the Weibel instability \cite{Weibel59} in spatially localized ion density accumulations such as shocks \cite{Stockem14b} and rarefaction waves \cite{Thaury10,Quinn12}. The magnetic field structures observed at later times expanded into the upstream region and they were not showing spatial oscillations on an electron skin depth-scale, which are typical for the magnetic fields driven by the Weibel instability. The magnetic fields were coherent over tens of electron skin depths and the area they covered was limited by the dimensions of the simulation box and by the simulation time. The large-scale magnetic fields started to grow when the shock normal was no longer aligned with the plasma flow velocity vector. We have attributed the large scale magnetic field to currents, which initially develop close to the cusp in the overlap layer. The simulation shows that the magnetic field eventually diffuses out of the overlap layer.

Experimental observations indicate that some SNR shocks are immersed in magnetic fields with amplitudes that exceed by far the values one would expect from the shock compression of the magnetic field of the interstellar medium \cite{Berezhko03}. Cosmic rays can magnetize the interstellar medium on large spatial scales \cite{Bell04}. We scale the growth time of the magnetic field and the size of the magnetic patches to the plasma parameters found close to SNR shocks in order to determine if the corrugation of plasma shocks could be important for the magnetic field generation at SNR shocks. We take the reference value 10 cm$^{-3}$ for the ion number density close to an SNR shock. Our simulation duration would correspond to $\approx 4\times 10^{-2}$ s. The spatial size $\approx 40\lambda_s$ of the magnetic field patches would correspond to about 50 km and their field amplitude would be of the order of 10 nT. The values for the growth time and the size of the magnetic field patches are microscopic compared to the size and the evolution time of an SNR shock. However, the magnetic field amplitude generated in our simulation exceeds that of the interstellar magnetic field by an order of magnitude. A corrugated shock front could thus generate magnetic fields ahead of the shock which are significantly stronger than those of the interstellar medium and it could compress these as it propagates across them. 

The periodic boundary conditions along the y-direction have limited the lateral expansion of the magnetic field patch at late times. An electron, which moves at the thermal speed, could cross the simulation box several times along the y-direction during the simulation time. Numerical artifacts, which are caused by the wrap-around of electrons, can usually be neglected because the electrons are scattered on the way by the electrostatic simulation noise. 

The key findings of this paper should however not be affected by the periodic boundary conditions. The net current that drives the magnetic field is generated in a small spatial interval close to the cusps that is far from the boundaries. The simulation box geometry will affect the shape of the generated magnetic field, but not its generation mechanism. The stability of the shocks is also not affected by the boundary conditions because the thermal speed of the ions is not high enough to let them cross the simulation box during the simulation time.

{\bf Acknowledgements:} The simulations were performed on resources provided by the Swedish National Infrastructure for Computing (SNIC) at HPC2N (Ume\aa ). GS acknowledges the EPSRC grant EP/L013975/1. The EPOCH code used in this research was developed under UK Engineering and Physics Sciences Research Council grants EP/G054940/1, EP/G055165/1 and EP/G056803/1.



\begin{thebibliography}{100}

\bibitem{Bale05} S. D. Bale, M. A. Balikhin, T. S. Horbury, V. V. Krasnoselskikh, H. Kucharek, E. M\"obius, S. N. Walker, A. Balogh, D. Burgess, B. Lembege, E. A. Lucek, M. Scholer, S. J. Schwartz, and M. F. Thomsen, Space Sci. Rev. \textbf{118} 161 (2005).
\bibitem{Burgess05} D. Burgess, E. A. Lucek, M. Scholer, S. D. Bale, M. A. Balikhin, A. Balogh, T. S. Horbury, V. V. Krasnoselskikh, H. Kucharek, B. Lembege, E. M\"obius, S. J. Schwarz, M. F. Thomsen, and S. N. Walker, Space Sci. Rev. \textbf{118} 205 (2005).
\bibitem{Dieckmann14} M. E. Dieckmann, G. Sarri, D. Doria, H. Ahmed, and M. Borghesi, New J. Phys. \textbf{16} 073001 (2014).
\bibitem{Kazimura98} Y. Kazimura, F. Califano, J. I. Sakai, T. Neubert, F. Pegoraro, and S. Bulanov, J. Phys. Soc. Jpn. \textbf{67} 1079 (1998).
\bibitem{Frederiksen04} J. T. Frederiksen, C. B. Hededal, T. Haugbolle, and A. Nordlund, Astrophys. J. \textbf{608} L13 (2004).
\bibitem{Spitkovsky08} A. Spitkovsky, Astrophys. J. \textbf{673} L39 (2008).
\bibitem{Stockem14a} A. Stockem, F. Fiuza, A. Bret, R. A. Fonseca, and L. O. Silva, Sci. Rep. \textbf{4} 3934 (2014).
\bibitem{Koopman67} D. W. Koopman, and D. A. Tidman, Phys. Rev. Lett. \textbf{18} 533 (1967).
\bibitem{Dean71} S. O. Dean, E. A. McLean, J. A. Stamper, and H. R. Griem, Phys. Rev. Lett. \textbf{27} 487 (1971).
\bibitem{Bell88} A. R. Bell, P. Choi, A. E. Dangor, O. Willi, and D. A. Bassett, Phys. Rev. A \textbf{38} 1363 (1988).
\bibitem{Romagnani08} L. Romagnani, S. V. Bulanov, M. Borghesi, P. Audebert, J. C. Gauthier, K. L\"owenbr\"uck, A. J. MacKinnon, P. Patel, G. Pretzler, T. Toncian, and O. Willi, Phys. Rev. Lett. \textbf{101} 025004 (2008).
\bibitem{Gregori12} G. Gregori \textit{et al.}, Nat. \textbf{481} 480 (2012).
\bibitem{Ahmed13} H. Ahmed, M. E. Dieckmann, L. Romagnani, D. Doria, G. Sarri, M. Cerchez, E. Ianni, I. Kourakis, A. L. Giesecke, M. Notley, R. Prasad, K. Quinn, O. Willi, and M. Borghesi, Phys. Rev. Lett. \textbf{110} 205001 (2013).
\bibitem{Hershkowitz81} N. Hershkowitz, J. Geophys. Res. \textbf{86} 3307 (1981).
\bibitem{Karimabadi91} H. Karimabadi, N. Omidi, and K. B. Quest, Geophys. Res. Lett. \textbf{18} 1813 (1991).
\bibitem{Kato10} T. N. Kato, and H. Takabe, Phys. Plasmas \textbf{17} 032114 (2010).
\bibitem{Forslund70} D. W. Forslund, and C. R. Shonk, Phys. Rev. Lett. \textbf{25} 281 (1970).
\bibitem{Dawson83} J. M. Dawson, Rev. Mod. Phys. \textbf{55} 403 (1983).
\bibitem{Dupree63} T. H. Dupree, Phys. Fluids \textbf{6} 1714 (1963).
\bibitem{Sarri11} G. Sarri, G. C. Murphy, M. E. Dieckmann, A. Bret, K. Quinn, I. Kourakis, M. Borghesi, L. O. C. Drury, and A. Ynnerman, New J. Phys. \textbf{13} 073023 (2011).
\bibitem{Eidmann00} K. Eidmann, J. Meyer-ter-Vehn, and T. Schlegel, Phys. Rev. E \textbf{62} 1202 (2000).
\bibitem{Stockem14b} A. Stockem, T. Grismayer, R. A. Fonseca, and L. O. Silva, Phys. Rev. Lett. \textbf{113} 105002 (2014).
\bibitem{Thaury10} C. Thaury, P. Mora, A. Heron, and J. C. Adam, Phys. Rev. E \textbf{82} 016408 (2010).
\bibitem{Quinn12} K. Quinn, L. Romagnani, B. Ramakrishna, G. Sarri, M. E. Dieckmann, P. A. Wilson, J. Fuchs, L. Lancia, A. Pipahl, T. Toncian, O. Willi, R. J. Clarke, M. Notley, A. Macchi, and M. Borghesi, Phys. Rev. Lett. \textbf{108} 135001 (2012).
\bibitem{Weibel59} E. S. Weibel, Phys. Rev. Lett. \textbf{2} 83 (1959).
\bibitem{Stockem09} A. Stockem, M. E. Dieckmann, and R. Schlickeiser, Plasma Phys. Controll. Fusion \textbf{51} 075014 (2009).
\bibitem{Forslund71} D. W. Forslund, and J. P. Freidberg, Phys. Rev. Lett. \textbf{27} 1189 (1971).
\bibitem{Dieckmann13} M. E. Dieckmann, H. Ahmed, G. Sarri, D. Doria, I. Kourakis, L. Romagnani, M. Pohl, and M. Borghesi, Phys. Plasmas \textbf{20} 042111 (2013).
\bibitem{Hansen06} J. F. Hansen, M. J. Edwards, D. H. Froula, A. D. Edens, G. Gregori, and T. Ditmire, Phys. Plasmas \textbf{13} 112101 (2006).

\bibitem{Berezhko03} E. G. Berezhko, L. T. Ksenofontov, and H. J. Volk, Astron. Astrophys. \textbf{412} L11 (2003).
\bibitem{Bell04} A. R. Bell, Monthly Not. R. Astron. Soc. \textbf{353} 550 (2004).
\end{thebibliography}

\end{document}